\documentclass[twocolumn,showpacs,preprintnumbers,amsmath,amssymb]{revtex4}

\usepackage{graphicx}
\usepackage{dcolumn}
\usepackage{bm}

\begin{document}
\title{An asymptotic model for the Kelvin-Helmholtz and Miles mechanisms of water wave generation by wind}
\author{Yuri M. Shtemler}
\email{shtemler@bgu.ac.il}
\author{Michael Mond}%
\email{mond@bgu.ac.il}
\author{Vladimir Cherniavskii}%
\email{vcherniavski@gmail.com}
\author{Ephim Golbraikh}
\email{golbref@bgu.ac.il}
\author{Yaakov Nissim}
\email{nasim@bgu.ac.il}
\altaffiliation{\\} \affiliation{*,\dag,\P,
Department of Mechanical Engineering,
\S Center for MHD studies,\\Ben-Gurion University of the Negev, \\
P.O.Box 653, Beer-Sheva 84105,
Israel,\\
%
%
\ddag  Institute of Mechanics \\  Moscow State University,\\
P.O.Box B-192, Moscow 119899, Russia}

\date{\today}

\begin{abstract}
The generalized Kelvin-Helmholtz (KH) and Miles mechanisms of the water wave generation by wind  are investigated for two-layer piece-wise linear model of the wind profile. It is shown by asymptotic expansions in small air-to-water density ratio that two mechanisms of the instability operate in quite different scales.
Miles' short waves are generated by weak winds, in particular, Miles' regime is responsible for initiation of the instability at the minimum wind speed, while the generalized KH regime dominates at strong winds and raises moderately-short waves.
\end{abstract}
\pacs{92.60.Cc, 92.10.Fj }
\maketitle
\section{Introduction}\label{sec:intro}
The current study is motivated by the attempt to understand and explain qualitatively water wave generation by winds. In particular, it aims at shedding light on and gaining a deeper understanding of the relative roles played by the Kelvin-Helmholtz (KH) and  Miles mechanisms in various regimes of wavelength and wind-strength values.

The Kelvin-Helmholtz and Miles theories are traditionally two principal models which describe the energy and momentum transfer from wind to waves. The classical KH instability (KHI) theory considers a tangential discontinuity between uniform flows of air and water \cite{Drazin 2002}.
Exact solutions of the linear KHI can be obtained due to the constant profiles of the unperturbed velocities of both air and water. Consequently, the growth rate of the KH mode is evaluated explicitly. It predicts a minimum wind speed necessary to make the waves grow against the stabilizing effects of gravity and surface tension. The quasi-laminar Miles model takes turbulence effects into account through
the mean profile of the unperturbed wind which  continuously varies  with the height above the air-water interface \cite{Miles 1957}.
The characteristic feature of the  Miles model is the concept of resonant wind-wave interaction, which assumes an intensive exchange of energy and momentum between the collective wave motion and the air particles within the critical layer (a narrow vicinity of the point within the air flow where the perturbation phase velocity is equal to the wind speed).

In the limit of small air-to-water density ratio $\epsilon^2=\rho_a/\rho_w$, Miles' model yields the following threshold values of the minimum wind speed and corresponding wavelength for the onset of surface waves: $U^{(cr)}_a\approx\sqrt{2g L_c}\approx0.23m/s$,
$\lambda^{(cr)}=2\pi L_c\approx 1.7\cdot 10^{-2}m$
($L_c=\sqrt{\sigma/(\rho_w g)}$ is the capillary length;  $\sigma$,  $g$, $\rho_a$  and $\rho_w$   are the air-water surface tension, gravity acceleration, air and water density, respectively;  $\sigma/\rho_w\approx 7.2\cdot 10^{-5}m^3/s^2$, $L_c\approx2.7\cdot 10^{-3}m$).
In the leading order in  $\epsilon$ the corresponding values for the KHI are  $U^{(cr)}_a\approx\epsilon^{-1}\sqrt{2g L_c}\approx7m/s$,
$\lambda^{(cr)}=2\pi L_c\approx 1.7\cdot 10^{-2}m$.
 The same value of the critical wave lengths ($\lambda^{(cr)}=2\pi L_c$) in both models at quite different
threshold velocities ($U^{(cr)}_a\sim\epsilon^0\sqrt{2g L_c}$ and $U^{(cr)}_a\sim\epsilon^{-1}\sqrt{2g L_c}$)
 sometimes leads to the misconception of the Miles and KH mechanisms of surface wave instability as alternative and competing mechanisms of wave generation in the same wave-length range. Thus, according to a widespread opinion, the Miles model is an improvement of the KH model (e.g. \cite{Barnett Kenyon 1975}),
 and the KH regime has no physical meaning in the context of the surface wave generation.
  Such a point of view may be traced to \cite{ Miles 1957} where it is written that 'The model to be developed here improves on the Kelvin-Helmholtz model by allowing for distributed, rather than concentrated vorticity...',
  in spite of the quite definite point of view later formulated  in \cite{Miles 1959}: '...the KH mechanism of instability still should be physically significant, albeit not responsible for the initial formation of water waves.'
An expression of such a belief may be found in \cite{Alexakis et al 2002} who have artificially superposed a velocity jump at the water-air interface on a continuous wind profile. That has been done in order to allow the excitation of the KHI at short wavelengths range similar to that of the Miles mechanism, namely, at the order of the capillary length. In fact, the unphysical negative phase velocity of the KH mode, exhibited in  \cite{Alexakis et al 2002} for a smooth wind profile, is a direct result of such artificially introduced concentrated vorticity. As is shown in the current work,  the KHI does occur for continuous profile but for longer wavelengths and higher wind speeds than in Miles' regime, and  the different scaling of both the characteristic wavelength and wind speed with $\epsilon$ provides an important criterion to distinguish between the generalized KH and  Miles mechanisms. Note that for wave lengths much larger than the capillary length, the classical KH model yields for the threshold wind speed  $U^{(cr)}_a\approx\epsilon^{-1}\sqrt{g /k}$, a value that depends on the wave length ($k=2\pi/\lambda$).
This leads to the proper characteristic wind speed for the KH-like regime $U^{(cr)}_a\sim\epsilon^{-3/2}\sqrt{g L_c}$ which is achieved at  wave lengths of the order of  $\lambda\sim \epsilon^{-1}L_c$ (see below Section 4, where it was established that accounting for the surface tension effect  for prediction of the minimum wind speed, which is correct for Miles' regime, is asymptotically small at KH scales of wave lengths, and
consequently has no physical meaning in the classical KH model.

   Although various aspects of the role of the KHI in water wave generation have been discussed previously
   \cite{Miles 1959}-\cite{Gertsenshtein et al 1988},
    and conditions necessary for the dominant role of the KHI  have been well recognized ($U_a\gg \sqrt{gL_a}$  with $\lambda\gg L_a$, where $L_a$  is the characteristic length of the wind profile inhomogeneity, the roughness length, which is typically of $\sim L_c$) \cite{Morland  Saffman 1993}, the  dominant role of the Miles mechanism is sometimes expanded to relatively long waves ($\lambda\gg L_a$) and strong winds ($U_a\gg\sqrt{g L_a}$). In fact, as is shown in the current work
the KH and Miles mechanisms act in quite different scales of both characteristic wind speeds and wave lengths  (see Sections 3 and 4, where the corresponding orders in $\epsilon$ have been established). One should expect the KH model with a concentrated vorticity to be more satisfactory for strong wind speeds and perturbations with wavelengths that are much larger than the roughness length of the smooth profiles. In contrast, the Miles model is more adequate for prediction of stability characteristics for low wind speeds and wave lengths.
Finally, in addition to the large difference in the scales of the perturbation wavelength and wind velocity for the two modes, the growth rate of the KH mode is much larger than that for Miles' mode (see below Sections 3 and 4).

Although the numerical solution of the problem that describes the wave generation by a wind of any smooth profile may be easily developed for a wide range of parameters (see e.g. \cite{Alexakis et al 2002}), the simple  piece-wise linear (PWL) models with low number of the layers $N$ are widely used in geophysics and astrophysics 
\cite{Ruden 2002} -\cite{Shtemler et al 2008}.  In particular,  the two-layer PWL approximation for the wind speed profile is widely employed in order to simplify the study of the wind instability problem and simultaneously to conserve the main features of the wind-shear effects (see e.g. \cite{Caponi et al 1992}, \cite{Chan 1977}-\cite{Fabrikant  Stepanayantz 1998}).
The advantage of the  PWL approximations with low $N$ is in the explicit solutions for the eigenfunctions  which 
results in a polynomial dispersion relation between perturbation frequency and wave number, and
allow to gain deeper understanding of the mechanisms that govern the instability, and the corresponding principal scales for
wavelengths and wind speeds, magnetic fields etc.
 In general, such a description should be considered as the simplest schematic modeling of the different shear effects rather than an approximation to any smooth wind profile that may be justified by the uncertainty in our knowledge of the wind profile near the waved interface. The validity of two-layer PWL approximation of the smooth wind profiles will be discussed in the last section of the present study, while
the convergence of the results of the general PWL (Rayleigh) method with increasing  $N$ to those of the corresponding smooth profiles of the wind has been established in \cite{Gertsenshtein 1970}-\cite{Zang 2004}.

Simultaneous study of the the Miles and KH mechanisms of instability is complicated due to the necessity to solve the problem in two quite different  regimes (with different wavelength of the perturbations and the unperturbed wind speed). For instance, both \cite{Caponi et al 1992} and \cite{Fabrikant  Stepanayantz 1998} apply the same two-layer PWL approximation for the wind profile to different regimes.
Fabrikant and Stepanayantz \cite{Fabrikant  Stepanayantz 1998} investigate the resonant Miles mechanism of instability, which is characterized by a near resonance distribution of the growth-rate coefficient localized in the vicinity of two (for $\sigma>0$) resonant wave numbers.
The model by Fabrikant and Stepanayantz in the Miles' region of parameters (low wave length and wind velocity) yields numerically the critical parameters for instability that are in fair agreement with Miles' result: $U^{(cr)}_a\approx0.27m/s$  for wavelength $\lambda^{(cr)}\approx 1.7\cdot 10^{-2}m$, while the KH mode was out of their consideration.
However, the growth rate coefficient of the Miles mode in the two-layer PWL approximation is significantly larger than that for the smooth profiles in \cite{Alexakis et al 2002}. As distinct from \cite{Fabrikant  Stepanayantz 1998}, Capponi et al. \cite{Caponi et al 1992} have solved the problem in the KH parameter regime that has no-resonance nature. Their model yields that the instability starts from vanishing wind velocity and wave length (in the scales of the KH mode of relatively long waves and strong winds). Two marginal curves appear instead of one in the classical KH model, when the order $n$ of the dispersion relation increases from  $n=2$ in the classical piece-wise constant KH approximation (vortex sheet instability) to $n=3$  in the two-layer PWL KH-like approximation. One of the marginal stability curves corresponds to a generalized KH mode modified by the wind-shear effects, while the other marginal stability curve is completely induced by wind-shear effects on the KH-like mode. That shear-induced instability has been erroneously interpreted in \cite{Caponi et al 1992} as the Miles mode (see Section 4 of the present study).

The present study of the surface wave instability is carried out by asymptotic expansions in small air-to-water density ratio and applied to two-layer PWL wind approximation. The application of the asymptotic expansions to the simplified PWL model for the wind profile leads to explicit solutions for both the KH regime which takes into account shear effects and Miles' regime. The present study demonstrates that both mechanisms are responsible for wind instability, but the Miles and KH-like regimes are distinguished by the scales in the wavelengths and  wind speeds.

The paper is organized as follows. The dispersion relation for two-layer PWL approximation of the wind speed profile is presented in the next Section. The Miles  and  KH-like mechanisms of wave generation  are analyzed asymptotically in the small density ratio  in Sections 3 and 4, respectively. Summary and Discussion are presented in Section 5. Asymptotic relations for the Miles regime are derived in Appendix A.

\section{The dispersion relation for two-layer PWL profiles of wind speed }\label{sec:dispersion}
The unperturbed state is assumed to be in hydrostatic equilibrium such that
 $\nabla(P_j+\rho_j gy)=0$, where $P_j$  and $\rho_j$  are the pressure and density, and  $j=a,w$ stand for air and water. The equilibrium velocity is approximated by a PWL profile, in which the liquid is at rest (so that the effect of the water drift velocity is neglected):
\begin{eqnarray}
   \,\,\, \rho =\rho_w,    U=0 \,\,\,\, \,\mbox{for} \,\,\, \,  y\leq0,
 \,\,\,\,\,\,\,\,\,\,\,\,\,\,
 \nonumber
\\
 \rho =\rho_a,    U=S_a y \,\,\,
\,\,\,\, \,\mbox{for} \,\,\,  0<y<L_a,
 \nonumber
\\
 \rho =\rho_a,    U=U_a   \,\,\,\,\,\, \,\mbox{for} \,\,\, \,
  y\geq L_a, \,\,\,\,\,\,\,\,\,\,\,
\end{eqnarray}
where $U_a$  is the wind speed at the reference height $y=L_a$  in the air;
$L_a$  is the thicknesses of the shear layer in the air (that has the meaning of the roughness of the air-water interface);  $S_a$ is the wind shear.

The stability properties of the equilibrium state (1) are investigated by considering the following normal modes:
\begin{eqnarray}
\Phi(x,y,t)=\phi(y)exp(-i\omega t+ikx),
\end{eqnarray}
where $\Phi$  stands for any of the perturbed variables and  $\phi$   is its amplitude;   $\omega=\omega_r+i\omega_i$ and $C=\omega/k=C_r+iC_i$  are the complex frequency and phase velocity,  and  $k$ is a real wave number.

In the case of the  PWL approximation for the velocity in Eqs. (1), the  dispersion relation rewritten in the present notations is as follows (\cite{Caponi et al 1992}, \cite{Fabrikant  Stepanayantz 1998}):
\begin{eqnarray}
[\omega^2(\rho_w+\rho_a)-\rho_a \omega S_a
-(\rho_w-\rho_a)gk-\sigma k^3]\times
\nonumber
\\
\times[\omega-S_a F(kL_a)]=-\rho_a\omega^2 S_a [G(kL_a)+1],
\end{eqnarray}
where
\begin{equation}
F(x)]=\frac{1}{2}exp(-2x)-\frac{1}{2}+x,\,\,\,G(x)=exp(-2x)-1,
\end{equation}
$$
F(x)=x^2+O(x^3),\,\,\,G(x)=x+O(x^2)\,\,\mbox{for} \,\,x\ll1,
$$
$$
F(x)=x+O(x^0),\,\,\,G(x)=-1+O(exp)\,\,\mbox{for} \,\,x\gg1,
$$
and $O(exp)$  denote exponentially small values in $x\gg1$.

Gaining a deeper understanding of the nature of the Miles and KH regimes of instability is facilitated by the introduction of the following two sets of characteristic scales (denoted by stars):
\begin{eqnarray}
(i\,)\,\,\,\,\rho_*=\rho_w,\,\,\,\,L_*=L_a,\,\,\,\,t_*=\sqrt{L_*/g},\,\,\,\,
\\
(ii)\,\,\,\,\rho_*=\rho_w,\,\,\,\,L_*=L_g,\,\,\,\,t_*=\sqrt{L_*/g}.\,\,\,\,
\end{eqnarray}
where $L_g=U_a^2/g$ is the gravitational length. As the characteristic properties of the two modes of instability will be unfolded in the next sections, it will become clearer that the Miles mechanism is better described by scaling the physical variables with the characteristic values given in Eqs. (5), while the scales depicted in Eqs. (6) are naturally suited to describe the KH mechanism. For now, though, the physical variables are scaled in a general way as follows:
\begin{eqnarray}
\bar{k}=kL_*,\,\,\,\bar{\omega}=\omega t_*,\,\,\,\bar{\sigma}=\frac{\sigma}{g\rho_* L_*^2},\,\,\,
\nonumber
\\
\bar{C}=\frac{Ct_*}{L_*},\,\,\,\bar{U}=\frac{Ut_*}{L_*},\,\,\epsilon^2=\frac{\rho_a}{\rho_*}.
\end{eqnarray}

Consequently, the dispersion relation (3) is given now in the following dimensionless form
\begin{eqnarray}
[\bar{\omega}^2(1+\epsilon^2)-\epsilon^2\bar{\omega} \bar{S}_a
-(1-\epsilon^2)\bar{k}-\bar{\sigma} \bar{k}^3]\times
\nonumber
\\
\times[\bar{\omega}-\bar{S}_a F(\bar{k}\bar{L}_a)]=-\epsilon^2\bar{\omega}^2 \bar{S}_a [G(\bar{k}\bar{L}_a)+1],
\end{eqnarray}
and each of the particular cases (represented in Eqs. (5) or (6) is determined by fixing the values $L_a$  or $U_a$, respectively,
\begin{eqnarray}
(i)\,\bar{L}_a=1,\,\,\bar{U}_a=\frac{U_a}{\sqrt{gL_a}},\,\,\bar{S}_a=\bar{U}_a,\,
\bar{\omega}=\bar{\omega}_a,\,\bar{k}=\bar{k}_a,
\\
(ii)\,\,\,\bar{L}_a=\frac{gL_a}{U_a^2},\,\,\bar{U}_a=1,\,\,\bar{S}_a=\frac{1}{L_a},\,
\bar{\omega}=\bar{\omega}_g,\,\bar{k}=\bar{k}_g.
\end{eqnarray}
Thus, the system's stability is determined by the following three dimensionless parameters: density ratio $\epsilon^2$, surface tension $\bar{\sigma}$, and Froude number $Fr=U_a/\sqrt{gL_a}$  ($\bar{U}_a=Fr$ and  $\bar{L}_a=1/Fr^2$
in Eqs. (9) and (10), respectively).

Before turning to the solutions of the dispersion equation (8) it is noted that the dimensionless surface tension coefficient may be expressed as a ratio of two characteristic lengths:
 \begin{equation}
 \bar{\sigma}=L_c^2/L_*^2,
\end{equation}
where $L_c=\sqrt{\sigma/(g\rho_w)}$  is the constant capillary length. Doing so it is evident that the dimensionless surface tension coefficient is uniquely determined by the characteristic scale $L_*$ and vice versa.

Dispersion relation (8) is a cubic equation for $\bar{\omega}_a$  as a function of  $\bar{k}_a$, which may be solved analytically by employing the Cardano solution. However, such a solution results in complex expressions that conceal the physical meanings and tendencies of the various regimes of instability. Instead, in the following sections dispersion relation (8)  will be solved in various regimes of parameter space with the aid of an asymptotic expansion in the natural small parameter of the problem, i.e., $\epsilon^2=\rho_a/\rho_w\approx  10^{-3}$. As will be seen, such expansions result in simple and physically transparent solutions.

\section{Miles' regime: short waves generated by weak winds
 }\label{sec:Miles}
{\it Dispersion relation  in the leading order in $\epsilon$.---}
In order to investigate the Miles regime the dispersion relation (8) is investigated in terms of the characteristic scales  in Eqs. (5) denoting the corresponding dimensionless values of parameters by subscript $a$. To simplify the resulting relations, it is convenient to introduce the following auxiliary scaled variables:
 \begin{equation}
\tilde{\omega}=\bar{\omega}_a/\bar{U}_a,\,\,\,\,\tilde{C}=\bar{C}_a/\bar{U}_a.
\end{equation}
Substituting definition (12) into the dispersion relation (8) yields
\begin{eqnarray}
[\tilde{\omega}^2
-\frac{\bar{k}_a+\bar{\sigma} \bar{k}_a^3}{\bar{U}_a^2}][\tilde{\omega}-F(\bar{k}_a)]=
\nonumber
\\
-\epsilon^2
[\tilde{\omega}^3+\frac{\bar{k}_a}{\bar{U}_a^2}\tilde{\omega}
-F(\bar{k}_a)(\tilde{\omega}^2 -\tilde{\omega}+\frac{\bar{k}_a}{\bar{U}_a^2})
+\tilde{\omega}^2 G(\bar{k}_a)].
\end{eqnarray}

First, the generation of short waves (with wave length $\lambda$  of the order of roughness $L_a$) is considered, such that all the relevant physical variables are of the order of $\epsilon^0$  with respect to the small density ratio
\begin{eqnarray}
\bar{k}_a\sim\epsilon^0,\,\,\,\bar{U}_a\sim\epsilon^0,\,\,\,\bar{\sigma}\sim\epsilon^0,
\nonumber
\\
\tilde{\omega}\sim \bar{\omega}_a\sim\epsilon^0,\,\,\,
\tilde{C}\sim \bar{C}_a\sim\epsilon^0.
\end{eqnarray}

In that case, Eq. (13) has the following two solutions in leading order in
 $\epsilon$
\begin{equation}
\tilde{\omega}-F(\bar{k}_a)=0,\,\,\,\,\,\,
\tilde{\omega}^2-\frac{\bar{k}_a+\bar{\sigma} \bar{k}_a^3}{\bar{U}_a^2}=0.
\end{equation}
To classify the waves that are described in Eq. (15), it is useful to return momentarily to the dimensional variables in Eqs. (15):
\begin{equation}
\omega-S_aF(kL_a)=0,\,\,\,\,\,\,
\rho_w\omega^2-\rho_wgk-\sigma k^3=0.
\end{equation}
The second equation in (16) evidently describes the gravity-capillary waves within the current approximation of low density ratio, while the first equation in (16) represents the inertial waves $\omega-kU_a \equiv k(C-U_a)=0$  in the limit $\bar{k}_a=kL_a\to\infty$ ($F(\bar{k})\approx\bar{k}_a$ for $\bar{k}_a\to\infty$). With this in mind, these waves will be called inertial and gravity-capillary waves, respectively.

Both modes of wave propagation that are described in Eqs. (15) are stable to leading order. Furthermore, since all the coefficients in the dispersion relation (13) are real, the higher order corrections to   $\tilde{\omega}$ in Eqs. (15) are of  orders  $\epsilon^{2n} , n=1,2...$,  and are real. That line of reasoning breaks down, though, when the inertial wave merges with one of the gravity-capillary waves (which happens when  two roots of Eqs. (15) are merged into one  double root, see Appendix A). In that case, higher order corrections are of order of $\epsilon^n,n=1,2...$,  and may lead to instability. The condition for a double root to occur is:
\begin{equation}
\bar{U}_a=
\frac{\sqrt{\bar{k}_a+\bar{\sigma} \bar{k}_a^3}}{F(\bar{k}_a)},
\end{equation}
where the resonance condition is resolved with respect to $\bar{U}_a$. Thus the system can be driven to unstable state via a resonant interaction of the gravity-capillary wave on the water surface  with the inertial air wave, i.e. the transition from stable to unstable waves is associated with the merging of the two waves. The image of the resonance condition (17) in the Cardano solution of a cubic equation is the vanishing in leading order of the discriminant of the cubic equation with real coefficients. As a result, in such a case small corrections of the order of  $\epsilon^{2}$ neglected in the solutions (15) can provide two complex conjunct roots and, hence, to destabilize the system. The real part of the double roots of Eq. (13) to leading order in  $\epsilon$ are
\begin{equation}
\tilde{\omega}_r= F(\bar{k}_a),\,\,\,\tilde{C}_r= F(\bar{k}_a)/\bar{k}_a,\,\,\,
\end{equation}
while the growth rate to leading order  is (Appendix A)
\begin{equation}
\tilde{\omega}_i=\epsilon \sqrt{ \frac{1}{2} F(\bar{k}_a)[G(\bar{k}_a)+1]}.\,\,\,
\end{equation}

The instability that occurs due to the resonance condition (17) in flows with the two-layer PWL profile of wind speed is quite similar to Miles' resonance condition for the instability of flows with smooth velocity profiles (as is discussed in \cite{Fabrikant Stepanayantz 1998}), and  the corresponding unstable mode is called Miles mode. The present asymptotic approach leads to the exact resonance condition (17) which provides an explicit expression for the resonant wind speed that leads to instability. The resonant nature of the instability  is confirmed by the numerical solution of the same dispersion relation   by obtaining a smoothed near resonance solution  in an asymptotically small  $\epsilon$-vicinity of the resonant wave numbers \cite{Fabrikant Stepanayantz 1998}.

Although the scaled values of the complex frequency $\tilde{\omega}$  and phase velocity  $\tilde{C}$, are independent of the surface tension, the unscaled  complex frequency $\bar{\omega}_a=\tilde{\omega}\bar{U}_a$  and phase velocity $\bar{C}_a=\tilde{C}\bar{U}_a$   do depend on the surface tension  through  $\bar{U}_a$ in Eq.(17), which strongly varies with  $\bar{\sigma}$.  The unscaled dimensionless frequency, phase velocity and growth rate are given by:
\begin{eqnarray}
\bar{\omega}_{ar}=\sqrt{\bar{k}_a+\bar{\sigma} \bar{k}_a^3},\,\,\,
\nonumber
\\
\,\bar{C}_{ar}= \sqrt{\frac{1+\bar{\sigma} \bar{k}_a^2}{\bar{k}_a}},\,\,\,
\bar{\omega}_{ai}=\epsilon
\sqrt{\frac{G(\bar{k}_a)+1}{2 F(\bar{k}_a)}}.\,\,\,
\end{eqnarray}

\begin{figure}[!h]
\includegraphics[scale=0.45]{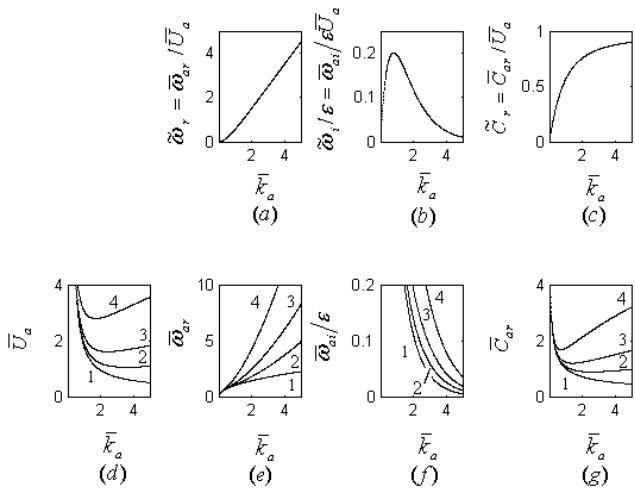}
\caption{\label{fig:epsfig1}
Dimensionless scaled:\\
(a) frequency $\tilde{\omega}_r=\bar{\omega}_{ar}/\bar{U}_a$;(b) growth rate
 $\tilde{\omega}_i/\epsilon=\bar{\omega}_{ai}/(\epsilon\bar{U}_a)$;
(c) phase velocity $\tilde{C}_r=\bar{C}_{ar}/\bar{U}_a$;\\  Dimensionless unscaled: \\
 (d)wind velocity $\bar{U}_{a}$; (e) frequency $\bar{\omega}_{ar}$;
 (f) growth rate $\bar{\omega}_{ai}/\epsilon$; (g) phase velocity $\bar{C}_{ar}$
vs wave number $\bar{k}_a$;  the curves marked by 1, 2, 3, 4 correspond to
$\bar{\sigma}=0; 0.15; 0.5; 2.0$.
 }
 \end{figure}

The frequency, growth rate, phase velocity and the resonant wind velocity are depicted for several values of $\bar{\sigma}$ (or, equivalently, the characteristic roughness length $L_*=L_a$, see relation (11))  as functions of the wave number in Fig.~\ref{fig:epsfig1}.

According to Fig.~\ref{fig:epsfig1}c, the resonant wind velocity depends on the wave number, or, physically more correctly, the resonant wave number depends on the wind velocity. For a nonzero surface tension ($\bar{\sigma}>0$) Fig.~\ref{fig:epsfig1} demonstrates a threshold (critical) wind speed $\bar{U}_a^{(cr)}$  for $\bar{k}_a=\bar{k}_a^{(cr)}$  necessary to make the waves grow (see also Table 1). In particular, for any given supercritical wind and wave velocities the resonance condition is satisfied for two wave numbers   (Fig.~\ref{fig:epsfig1}d), which correspond to waves with different growth rates: longer waves (with smaller wave numbers) grow faster, since the growth rate  monotonically increases with the wave length (Fig.~\ref{fig:epsfig1}f). Similarly the minimum gravity-capillary wave speed
 $\bar{C}_{ar}^{(cr)'}$  is achieved at the wave number $\bar{k}_a=\bar{k}_a^{(cr)'}$  that differs from that for the wind speed   (see Figs. 1d and 1g)
\begin{equation}
\,\bar{C}_{ar}^{(cr)'}= \root 4 \of {4\bar{\sigma}},\,\,\,\,\bar{k}_a^{(cr)'}=1/\sqrt{\bar{\sigma}}.
\end{equation}
In addition, according to Fig.~\ref{fig:epsfig1} the value of the roughness (or, equivalently, dimensionless surface tension) strongly influences the marginal stability values of the wind velocity as well as the wave phase velocity, while the growth rate depends only slightly on the roughness (see Fig.~\ref{fig:epsfig1}f).

The critical values of the dimensionless as well as the corresponding dimensional parameters
presented in Table I for several values of the dimensionless surface tension   $\bar{\sigma}= 0.15; 0.5; 2.0$
have been calculated at minimum wind velocity ($\bar{k}_a=\bar{k}_a^{(cr)}$)  by using the following relations:
\begin{eqnarray}
L_a=L_c/\sqrt{\bar{\sigma}},\,U_a=\bar{U}_a\sqrt{gL_a},\,C_a=\bar{C}_a\sqrt{gL_a},
\nonumber
\\
k_a=\bar{k}_a/L_a,\,\,\,\,\lambda_a=2\pi/k_a,\,\,\,\,L_g=U_a^2/g.
\end{eqnarray}
\begin{table}
  \caption{System parameters at minimum wind velocity \\($\bar{k}_a=\bar{k}_a^{(cr)}$) for Miles'  regime in air-water system.}
  \label{tab:I}
  \begin{center}
  \begin{tabular}{ccccccccc}
    \hline
$\bar{\sigma}$ & $\bar{k}_a^{(cr)}$ & $\bar{U}_a^{(cr)}$ & $\bar{C}_a^{(cr)}$ & $L_a$ &  $\lambda_a^{(cr)}$ & $U_a^{(cr)}$ & $L_g^{(cr)}$ & $C_a^{(cr)}$
\\
 &       &      &     & $cm$   & $cm$    & $cm/s$      & $cm$ & $cm/s$
     \\
      \hline
       0.15 & 3.6 & 1.0 & 0.9  & 0.7 & 1.2  & 26 & 0.7 & 23.5 \\
       0.5  & 2.5 & 1.6 & 1.2  & 0.4 & 1.0 & 33 & 1.1 & 23.8 \\
       2.0  & 1.9 & 2.7 & 1.6  & 0.2 & 0.7 & 38 & 1.5 & 33.7\\
      \end{tabular}
  \end{center}
\end{table}

{\it Analysis of results in Table I.---}
The critical value of the wave length $\lambda \sim (0.7\div1.2)\cdot10^{-2}m$ is of the order of the wind profile inhomogeneity (roughness), $L_a \sim (0.2\div0.7)\cdot10^{-2}m$, as well as of the order of the gravity and capillary lengths  $L_g \sim (0.7\div1.5)\cdot10^{-2}m$, $L_c \sim 0.27\cdot10^{-2}m$:
\begin{equation}
\lambda \sim L_a \sim L_g \sim L_c.
\end{equation}
The closeness of $L_a $  and $L_g$  follows from the following estimation for $\bar{U}_a^{(cr)} \sim 0.75\div1.6\sim 1$
\begin{equation}
\bar{U}_a=\frac{U_a}{\sqrt{gL_a}}\equiv \sqrt{\frac{L_g}{L_a}}\sim 1.
\end{equation}
The values of the critical parameter in Table 1 are close to those obtained numerically by \cite{Fabrikant Stepanayantz 1998}  for two-layer PWL profiles in Eq. (1).

At this point it is natural to discuss the validity of the results obtained above for the PWL model for a  smooth wind velocity profile. Since the wavelength is of the order of the roughness value in the critical region of parameters $\lambda^{(cr)}\sim L_a$ (Table 1), one should not expect a model with discontinuous profiles to be compared  favorably  with the smooth profiles being modeled, it rather models schematically the influence of the wind-shear effects on the interface instability. Indeed, the growth rate coefficients in Fig. 1b significantly overestimate the values calculated numerically for smooth wind profiles in \cite{Alexakis et al 2002}. Nevertheless, the values of the wind velocity, roughness and wave length in Table 1 are in  fair agreement by the order of magnitudes with experimental data for low wind velocities \cite{Wu 1968} ($L_a \approx 0.3\cdot10^{-2}m$  at  $U_a \approx 0.23m/s$). The latter results are also consistent with the results of Miles' model for the smooth logarithmic profile in the limit of small density ratio: $U_a\approx\sqrt{gL_c}\approx 0.33m/s$, $\lambda^{(cr)}\approx 1.7\cdot10^{-2}m$. As was mentioned above, the two-layer PWL approximation   leads to the present resonance model, which is  quite similar to the Miles critical-layer resonance model for  smooth velocity profiles \cite{Fabrikant Stepanayantz 1998}.

{\it Region of asymptotic nonuniformity in small $\bar{k}_a\sim \epsilon$.---}
The solution presented above loses its validity for sufficiently long waves $\bar{\lambda}_a=\lambda/L_a$ (small $\bar{k}_a=2\pi/\bar{\lambda}_a$). The nonuniformity of the asymptotic solution is exhibited in the infinite values of the resonant wind velocity  $\bar{U}_a(\bar{k}_a)$, growth rate $\bar{\omega}_{ai}(\bar{k}_a)$, and phase velocity  $\bar{C}_{ar}(\bar{k}_a)$ at low values of $\bar{k}_a$ in Fig. 1. That resonant solution has been obtained by a formal asymptotic expansion in $\epsilon$  that has been applied to the cubic dispersion equation (13), with the scaled eigenvalue  $\tilde{\omega}$  expanded as follows (Appendix):
\begin{equation}
\tilde{\omega}=\tilde{\omega}_0+\epsilon\tilde{\omega}_1+\epsilon^2\tilde{\omega}_2+...
\end{equation}

Such an expansion is a direct result of the assumptions  expressed in estimations (14), that both the wave number as well as the wind velocity are of zeroth leading order in $\epsilon$. However, as the wave lengths and the wind velocity increase, the asymptotic expansion (25) loses its validity. Thus, turning now to small values of the wave number $\bar{k}_a\sim\epsilon$ (see Appendix A),  and inspecting again dispersion relation (8), results in the following scalings of the remaining  physical variables:
\begin{eqnarray}
\bar{\omega}_{ar}\sim \sqrt{\bar{k}_a}\sim\sqrt{\epsilon},
\bar{\omega}_{ai}\sim \frac{\epsilon}{\sqrt{\bar{k}_a}}\sim\sqrt{\epsilon},
\bar{U}_{a}\sim \frac{1}{\bar{k}_a^{3/2}}\sim\frac{1}{\epsilon^{3/2}},
\nonumber
\\
\bar{C}_{ar}\sim \frac{1}{\sqrt{\bar{k}_a}}\sim\frac{1}{\sqrt{\epsilon}},\,\,\,
\bar{C}_{ai}\sim \frac{\epsilon}{\bar{k}_a^{3/2}}\sim\frac{1}{\sqrt{\epsilon}}.\,\,\,
\end{eqnarray}
Thus the region of nonuniformity indeed corresponds to the longer waves with $\bar{k}_a\sim\epsilon$   and higher wind velocity
$\bar{U}_{a}\sim\epsilon^{-3/2}$  than it was assumed in estimations (14). To further develop the asymptotic solution in the region of nonuniformity, it is recalled that the classical KH mode (vortex sheet instability) has the following orders in terms of the characteristic length  $L_g$ ( e.g. \cite{Shtemler et al 2008}):
\begin{eqnarray}
\bar{k}_g=k L_g\sim\frac{1}{\epsilon^2},\,\,
\bar{\omega}_g= \omega \sqrt{\frac{L_g}{g}}\sim\frac{1}{\epsilon},
\nonumber
\\
\bar{C}_g=\frac{C_g}{gL_g}\sim \epsilon,\,\,
\bar{U}_g\equiv1.
\end{eqnarray}
Rewriting  relations (26) in terms of the characteristic length $L_g$  yields by using estimations (26)-(27):
\begin{eqnarray}
\bar{k}_a=\bar{k}_g\frac{L_a}{L_g}\sim\frac{1}{\epsilon^2}\frac{L_a}{L_g},\,\,
\bar{\omega}_a= \bar{\omega}_g\sqrt{\frac{L_a}{L_g}}\sim\frac{1}{\epsilon}\sqrt{\frac{L_a}{L_g}},
\nonumber
\\
\bar{C}_a=\bar{C}_g\sqrt{\frac{L_g}{L_a}}\sim \epsilon\sqrt{\frac{L_g}{L_a}},\,\,
\bar{U}_a=\bar{U}_g\sqrt{\frac{L_g}{L_a}}\sim \sqrt{\frac{L_g}{L_a}}.
\end{eqnarray}
Comparing Eqs. (27) and (28) results in the following estimate of the characteristic roughness   for which the resonant solution describes the classical KH mode:
\begin{equation}
\bar{L}_a=L_a/L_g\sim\epsilon^3.
\end{equation}
This is also consistent with the estimation (26) for $\bar{U}_a=U_a/\sqrt{gL_a}\sim \epsilon^{-3/2}$  (accounting for the definition of the scaled dimensionless roughness length $\bar{L}_a=L_ag/U_a^2$). Substituting now (29) into (28) leads indeed to the asymptotic scales already defined above in relations (26). In particular, according to estimates (26) the growth rate of those moderately-short waves is expected to be much larger than that (see Eqs. (20)) for the Miles short waves.

The term moderately-short-wave mode stems from the fact that its wave length is much larger than that corresponding to the scales of the interface roughness and, simultaneously, much less than the scale of gravity waves:
$$
L_a\ll\lambda\ll L_g
$$
or in dimensionless form using estimations (28)-(29):
\begin{equation}
1\ll\bar{\lambda}_a=\frac{\lambda}{L_a}\sim\frac{1}{\epsilon}\ll\frac{L_g}{L_a}\sim\frac{1}{\epsilon^3}.
\end{equation}

Finally note that the assumption (14) for the asymptotic orders of the values correspond to the outer solution in terms of the method of matched asymptotic expansions \cite{Van Dyke 1964}, while within the region of nonuniformity (26) or (27) of the corresponding outer asymptotic solution (Miles' mode) the inner solution should be constructed.

\section{KH regime: moderately-short waves generated by strong winds}
{\it Dispersion relation  in the leading order in $\epsilon$.---}
The asymptotic solution is now constructed in the inner region of small $\bar{k}_a\sim\epsilon$, where the outer  solution described by relations (17) and (20), is not uniformly valid. Following estimates (26) it is useful to introduce scaled variables of zeroth order in the inner region:
\begin{eqnarray}
\hat{k}_a=\bar{k}_a/\epsilon\sim\epsilon^0,\,\,
\hat{\omega}_a= \bar{\omega}_a/\sqrt{\epsilon}\sim\epsilon^0,\,\,
\nonumber
\\
\hat{C}_a=\bar{C}_a\sqrt{\epsilon}\sim\epsilon^0,\,\,
\hat{U}_a=\bar{U}_a\epsilon^{3/2}\sim\epsilon^0.
\end{eqnarray}

Substituting the inner variables (31) into  dispersion relation (8), and expanding $F(\bar{k}_a)$  and  $G(\bar{k}_a)$  in Taylor series in  $\bar{k}_a\ll1$ by using (4),   
  yields in the leading order in $\epsilon$ the following reduced dispersion relation:
\begin{equation}
\hat{\omega}_a^3-\hat{U}_a\hat{k}_a^2\hat{\omega}_a^2
+\hat{\omega}_a\hat{k}_a(\hat{U}_a^2\hat{k}_a-1)+\hat{U}_a\hat{k}_a^3=0.
\end{equation}
In order to estimate the effect of the surface tension the experimentally observed roughness value  $L_a\sim10^{-3}$ is adopted  for $25m/s<U_a<50m/s$ (Fig. 3b in \cite{Powell et al 2003}). Hence, since $\bar{\sigma}=L^2_c/L^2_a\sim10$ ($L_c\sim3\cdot10^{-3}$), the surface tension term in the dispersion relation (32) that is  $\sim \epsilon^2\bar{\sigma}$ has been neglected in comparison with the remaining  terms $\sim \epsilon^0$, and hence does not appear in Eq. (32). Consequently, accounting for the surface tension effect in the  classical criterion of the KHI has no physical meaning.

{\it Gravitational length as the characteristic scale.---}
For comparison convenience with the limit of vortex sheet instability (i.e., the classical KHI limit) the dispersion relation (32) is rewritten in terms of the characteristic scales  in Eqs. (6) based on the gravitational length $L_g=U^2_a/g$ and scaled by $\epsilon$:
\begin{equation}
\hat{\omega}_g^3-\hat{L}_a\hat{k}_g^2\hat{\omega}_g^2
+\hat{\omega}_g\hat{k}_g(\hat{k}_g-1)+\hat{L}_a\hat{k}_g^3=0,
\end{equation}
or, equivalently, in terms of a complex phase velocity
\begin{equation}
\hat{C}_g^3-\hat{L}_a\hat{k}_g\hat{C}_g^2
+\hat{C}_g(1-\hat{k}_g^{-1})+\hat{L}_a=0,
\end{equation}
where
\begin{eqnarray}
\hat{k}_g=\epsilon^2\bar{k}_g\sim\epsilon^0,\,\,
\hat{\omega}_g=\epsilon^2\bar{\omega}_g\sim\epsilon^0,\,\,
\nonumber
\\
\hat{C}_g=\epsilon^{-1}\bar{C}_g\sim\epsilon^0,\,\,
\hat{L}_a=\epsilon^{-3}\bar{L}_a\sim\epsilon^0.
\end{eqnarray}
For such a choice of the characteristic scales, the dispersion relation (33) or (34) explicitly contains the dimensionless roughness $\hat{L}_a=L_a/L_g$, so that the latter may be set to zero, and thus obtain  the dispersion relation for the classical KHI limit (see Eq. (39) below).

In Table II the  wave lengths $\lambda$   are estimated at  wind speeds $U_a=25$ and $50m/s$ ($L_a\sim 10^{-3}m$, \cite{Powell et al 2003}) for the wave numbers $\hat{k}_g=2\pi \epsilon^2U_a^2/(g\lambda)$ of the order of  $\epsilon^0$. It is seen that the typical wave lengths in the KH regime are ranged from $\lambda\sim 0.1m$  at $U_a=25m/s$  to  $\lambda\sim 10m$  at $U_a=50m/s$.

\begin{table}
  \caption{ Typical parameters of the air-water system at hurricane winds for KH mode.}
  \label{tab:II}
  \begin{center}
  \begin{tabular}{ccccccccc}
    \hline
$U_a$ & $\hat{k}_g$ & $\lambda=2\pi\epsilon^2\frac{U_a^2}{g\hat{k}_g}$ & $L_a$ & $\hat{L}_a=\frac{L_ag}{\epsilon^3U_a^2}$
\\
 $m/s$     &      &  $m$   & $m$   &    &
     \\
      \hline
       25  & 0.2 &  2.5  & 0.001 & 0.4  \\
       25  & 1.0 &  0.5  & 0.001 & 0.4  \\
       25  & 5.0 &  0.1  & 0.001 & 0.4 \\
       50  & 0.2 &  10.  & 0.001 & 0.1  \\
       50  & 1.0 &  2.0  & 0.001 & 0.1  \\
       50  & 5.0 &  0.4  & 0.001 & 0.1 \\
      \end{tabular}
  \end{center}
\end{table}
{\it General relations for marginal stability curves.---}
The cubic equation (34) with real coefficients has either one real and two complex conjugate roots, or three real roots at least two of which are equal, or three different real roots. The various cases depend on the value of the discriminant $Q$  that may be either positive or zero, or negative, respectively. It is convenient to write the condition of zero discriminant in the following way:
\begin{equation}
\alpha\hat{L_a}^4+\beta\hat{L_a}^2+\gamma=0,
\end{equation}
where
$$
\alpha(\hat{k}_g)=4\hat{k}_g^3,\,\beta(\hat{k}_g)=\hat{k}_g^2-20\hat{k}_g-8, \, \gamma(\hat{k}_g)=-4(1-\hat{k}_g^{-1})^3.
$$
Consequently, the corresponding (physically meaningful, i.e. only positive) marginal roughness lengths are
\begin{equation}
\hat{L}_a^{(1,2)}(\hat{k}_g)=\sqrt{\frac{-\beta\pm\sqrt{\beta^2-4\alpha\gamma}}{2\alpha}},
\,\,\,(\hat{L}_a^{(1)}<\hat{L}_a^{(2)}).
\end{equation}
In general, Eq. (36) has either two or one positive  solutions (37). The number of solutions depends on the existence of a real solution for $\hat{L}_a^{(1)}$, while the  instability occurs in one of the following two cases: the region of instability lies above either the marginal stability curve $\hat{L}_a=\hat{L}_a^{(1)}(\hat{k}_g)$ or the value $\hat{L}_a=0$, but below the marginal stability curve $\hat{L}_a=\hat{L}_a^{(2)}(\hat{k}_g)$:
$$
\hat{L}_a^{(1)}<\hat{L}_a<\hat{L}_a^{(2)}\,\, \mbox{or}\,\,0<\hat{L}_a<\hat{L}_a^{(2)}.
$$

Consider now the marginal stability  frequency $\hat{\omega}_{gm}$  that is obtained by setting the imaginary value of   $\hat{\omega}$ to zero ($\hat{\omega}_{gi}=0$). This may be equivalently obtained by setting the discriminant  $Q=0$ equal to zero. The marginal frequency and phase velocity are then given by:
\begin{equation}
\hat{\omega}_{gm}=\hat{k}_g\hat{C}_{gm},\,\,
\hat{C}_{gm}=-\frac{4}{3}\hat{k}_g^3+\root 3 \of {\frac{q}{2}},\,\,
\end{equation}
where
$$
\frac{q}{2}=\hat{L}_a\big(-\frac{\hat{L}_a^2\hat{k}_g^3}{27}
+\frac{\hat{k}_g-1}{6}+\frac{1}{2}\big).
$$

The dimensionless values of the characteristic roughness, frequency, growth rate, real and imaginary phase velocities vs the wave number are depicted in Figure 2. The marginal curves 1 and 2  set the boundaries of the instability region in Figs. 2a-2e. It will be shown briefly in the next subsection that the marginal stability curves 1 and 2 represent KH-like and inertial-like modes, respectively. In Figure 2a the marginal stability curve 1, $\hat{L}_a=\hat{L}_a^{(1)}(\hat{k}_g)$, corresponds to  the classical KH mode generalized by accounting for the wind-shear  effects.  The appearance of the second marginal curve 2, $\hat{L}_a=\hat{L}_a^{(2)}(\hat{k}_g)$, which is absent in the classical KH mode, is entirely due to the wind-shear  effects.  At constant values of the roughness its influence,  as seen from Figs. 2a and 2c,   leads to cut off of the growth rates in the short wave limit. Curves 3 and 4 depict the values of the complex frequency  and phase velocity  for $\hat{L}_a=0.5$  and  $\hat{L}_a=0.12$, which correspond to  $L_a\sim10^{-3}m$ ($U_a=25m/s$ and $U_a=50m/s$,  \cite{Powell et al 2003}).  In Figures 2c and 2e the growth-rate curves 5 (imaginary frequency and phase velocity)  are depicted for the classical  KH approximation  for the same wind speeds $U_a=25m/s$  and  $U_a=50m/s$. It is seen that they are very close to those for the PWL approximation in the whole range of the wave numbers at $U_a=50m/s$, but only at slightly supercritical wave numbers at $U_a=25m/s$.

\begin{figure}[!h]
\includegraphics[scale=0.45]{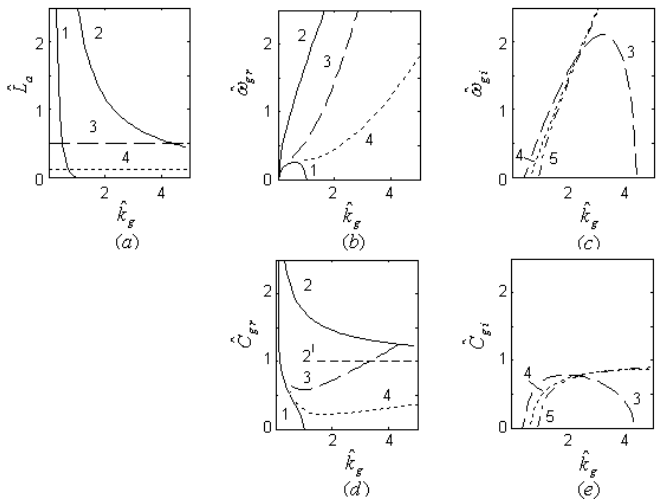}
\caption{\label{fig:epsfig2}
Marginal stability and growth rate curves vs wave number.
(a)	roughness $\hat{L}_a$; (b) frequency $\hat{\omega}_{gr}$; (c) growth rate $\hat{\omega}_{gi}$; (d) real phase velocity  $\hat{C}_{gr}$; (e) imaginary phase velocity  $\hat{C}_{gi}$.
the KH-like and inertial-like marginal curves 1 and 2 correspond to signs - and + in Eq. (37);
the straight line 2$'$ is the asymptote $\hat{C}_{gr}=1$  at $\hat{k}_g\to\infty$.
Dashed curves 3 and pointed curves 4 depict growth rates for  $\hat{L}_a=0.5$,  $\hat{L}_a=0.12$  at $U_a=25m/s$,  $U_a=50m/s$, respectively ($L_a=10^{-3}m$).
Point-dashed curves 5 in Figs. 2c and 2e depict growth rates
for the classical KH approximation $\hat{\omega}_{gi}=\sqrt{\hat{k}_g^2-\hat{k}_g}$.
 }
 \end{figure}

{\it KH and inertial limits for marginal curves.---}
The limit $\hat{L}_a\rightarrow 0$  corresponds to the classical KH mode (see Fig. 2a), and the point $\hat{k}_g=1$ where the marginal curve 1 crosses the  $\hat{k}_g$-axis is the threshold of the classical KHI, so that the instability occurs for  $\hat{k}_g>1$. Indeed,  the dispersion relation (33) may be additionally expanded in  $\hat{L}_a\ll1$, and thus reduced to the limit of the classical KH mode (vortex sheet instability):
\begin{equation}
\hat{\omega}_g^3+\hat{\omega}_g\hat{k}_g(\hat{k}_g-1)=0,
\end{equation}
where the trivial solution $\hat{\omega}_g=0$  should be omitted. To make the comparison evident the dispersion relation (39) may be rewritten in dimensional form as:
\begin{equation}
\omega^2+\frac{\rho_a}{\rho_w}U_a^2k^2-gk=0.
\end{equation}
Thus, a  vicinity of the  $\hat{k}_g$-axis in Fig. 2a is the region of applicability of the classical KH approximation, in particular, the region includes a part of the marginal curve 1, where $\hat{L}_a\ll1$, and  the marginal curves 1 are termed KH-like.

On the other hand, the condition  $\hat{L}_a\ll1$ is only necessary but not sufficient for the convergence of the KH-like theory to the classical KHI limit. Indeed, according to Eqs. (36) in the limit of  $\hat{k}_g\rightarrow\infty$ the roughness length is given by:
\begin{equation}
\hat{L}_a^{(2)}\approx 2/\hat{k}_g\rightarrow 0,
\end{equation}
while the dispersion relation (34) yields:
\begin{equation}
\hat{\omega}_g^3-2\hat{k}_g\hat{\omega}_g^2
+\hat{\omega}_g\hat{k}_g^2=0.
\end{equation}

Equation (42) is indeed quite different from Eq. (39) for the classical KH mode. The solution of Eq. (42) in the limit of  $\hat{k}_g\rightarrow\infty$ (represented by the dashed straight line $2'$, in Fig. 2d) is given by:
\begin{equation}
\hat{C}_g=1.
\end{equation}

Remembering now that the phase velocity has been made dimensionless by using the characteristic velocity $U_a$, equality (43) may be rewritten in dimensional form as  $C=U_a$. Hence, the corresponding marginal curves 2 in Fig. 2 are termed inertial-like.

{\it Definition  of the KH-like  regime of instability.---}
Following the discussion in the previous sections, the region between the marginal curves 1 and 2 in Fig. 2, is called the KH-like regime or finite-shear-KH regime.
%
Such a definition seems to be physically plausible and convenient in order to distinguish the KH-like mode that is
characterized by moderately-short waves and strong winds
 from  Miles' mode that is characterized by short waves and weak winds, as well as by the resonant nature of the instability.

Note that the high-$\hat{k}_g$  limit of the upper marginal curve 2 in Fig. 2d in the region of the KH-like instability is matched asymptotically with the axis  $\bar{C}_g=0$ for the Miles' mode at $\epsilon\rightarrow0$. To see that, according to the conventional asymptotic matching procedure of the inner and outer expansions \cite{Van Dyke 1964}, the inner (KH-like) marginal stability curve $\hat{C}_g=1$  in (43) should be rewritten in the outer (Miles) limit. Then by using relations (28) and (35) between the inner and outer variables $\hat{k}_g$, $\hat{C}_g$  and  $\bar{k}_a$,  $\bar{C}_a$, one has that the region of  the KH-like instability  shrinks in the outer Miles limit since the marginal curve 2 in Fig. 2 approaches zero in that limit ($\hat{k}_g\rightarrow\infty$, $\epsilon\rightarrow0$).

{\it Comparison    with the numerical results  in \cite{Caponi et al 1992}.---}
The results presented in Fig 2a have been reconstructed in the plane ($\sqrt{kL_g}$, $kL_a$) for comparison with the numerical results  presented in \cite{Caponi et al 1992}. The marginal curves obtained for the present asymptotic solutions in $\epsilon$ in Fig. 3, are very close to those calculated numerically (see Fig. 3 in \cite{Caponi et al 1992}). According to (41) the upper marginal curve  $\hat{L}_a^{(2)}$   vanishes as $2/\hat{k}_g$  for  $\hat{k}_g\rightarrow\infty$ ($\hat{k}_g\hat{L}_a^{(2)}\to2$). Using relations (28) and (35) this yields the constant asymptotic limit for the upper marginal curve $kL_a=2\epsilon$  for $kL_g=\infty$. The point of touching the bottom marginal curve  in Fig. 3 ($\sqrt{kL_g}=\epsilon^{-1}$, $kL_a=0$) corresponds to the point ($\hat{k}_g=1$, $\hat{L}_a=0$) on the bottom of the marginal curve in Fig. 2a. Thus, having in mind that the region characterized by low values of  $\hat{L}_a$, and $\hat{k}_g>1$  is the region of applicability of the classical KH mode, it may be concluded that the point is the critical point for the classical KH mode rather than the point of 'a continuous transition from the Miles-type instability to the KH-type instability' (cited from \cite{Caponi et al 1992}).

\begin{figure}[!h]
\includegraphics[scale=0.45]{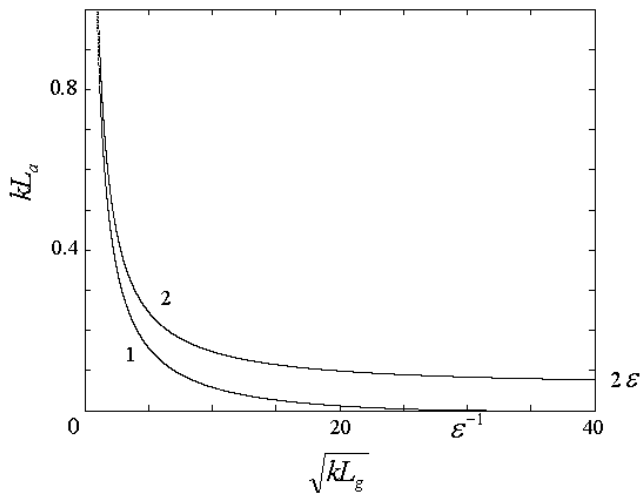}
\caption{\label{fig:epsfig3}
KH-like and inertial-like marginal curves 1 and 2,  $kL_a$  vs  $\sqrt{kL_g}$
(for comparison with the numerical results \cite{Caponi et al 1992} for zero water drift speed).
The instability region is between the marginal curves 1 and 2.
The asymptotic abscissa $\sqrt{kL_g}=\epsilon^{-1}$ of the touching point $kL_a=0$ of the bottom marginal curve
is the threshold point of the classical KHI ($\hat{k}_g=1$,$\hat{L}_a=0$);
$kL_a=2\epsilon$ is the asymptotic limit for the upper marginal curve for  $kL_g\rightarrow\infty$.
 }
 \end{figure}

 The closeness of the marginal stability curves obtained  numerically by \cite{Caponi et al 1992} for arbitrary  $\bar{L}_a$ to those in the present study, in spite of the fact that the latter has been selected on the basis  of the asymptotic scaling hypothesis (29) for the classical KH regime for small  $\bar{L}_a$, supports the result dispersion relation (33)

\section{Conclusions and discussion.}
Asymptotic expansion in the small   air-to-water mass densities ratio $\epsilon^2\ll1$ that has been applied to the two-layer PWL approximation of wind profile (the simplest modeling of wind-shear effects) provides an explicit and physically transparent solution of the dispersion relation. Based on the conventional method of matched asymptotic expansions \cite{Van Dyke 1964}, it has been shown that the Miles and KH regimes are described by the outer and inner asymptotic solutions in small $\epsilon$, respectively, and operate at quite different scales of the unperturbed wind speeds and perturbation wave lengths:

(i) The Miles mechanism is relevant for weak winds, which generate short waves with wavelengths  $\lambda$ of the order of the scales of the wind profile inhomogeneity (roughness) $L_a$, as well as the characteristic capillary and gravitational length, $L_c$  and  $L_g$ 
\begin{eqnarray}
\frac{\lambda}{L_a}\sim\frac{L_c}{L_a}\sim\frac{L_g}{L_a}\sim \epsilon^0,\,\,
Fr\sim \epsilon^0,\,\,
\frac{\sigma}{\rho_wg\lambda^2}\sim \epsilon^0,\,\,
\nonumber
\\
\omega_r \sqrt{\frac{L_a}{g}}\sim \epsilon^0,\,
\omega_i\sqrt{\frac{L_a}{g}}\sim \epsilon,\,
\frac{C_r}{\sqrt{L_a g}}\sim \epsilon^0,\,
\frac{C_i}{\sqrt{L_a g}}\sim \epsilon.
\end{eqnarray}
 (ii) The KH mechanism is relevant for strong winds which generate moderately-short waves with wavelengths  $\lambda$  much larger than the roughness $L_a$  and the capillary length  $L_c$, but much shorter than the gravitational length   $L_g$
($L_a\sim L_c\ll\lambda\ll L_g$)
\begin{eqnarray}
\frac{\lambda}{L_a}\sim\frac{1}{\epsilon},
\frac{L_c}{L_a}\sim\epsilon^0,
\frac{L_g}{L_a}\sim \frac{1}{\epsilon^3},
Fr\sim\frac{1}{\epsilon^{3/2}},
\frac{\sigma}{\rho_wg\lambda^2}\sim \epsilon^2,
\nonumber
\\
\omega_r \sqrt{\frac{L_a}{g}}\sim \omega_i\sqrt{\frac{L_a}{g}}\sim \epsilon^{1/2},\,
\frac{C_r}{\sqrt{L_a g}}\sim \frac{C_i}{\sqrt{L_a g}}\sim  \frac{1}{\epsilon^{1/2}}.
\end{eqnarray}

According to estimates (44) and (45) for  Froude numbers the characteristic wind speed in the KH-like regime,  $U_a\sim\epsilon^{-3/2}\sqrt{gL_a}$, is much larger than that, $U_a\sim\epsilon^{0}\sqrt{gL_a}$,   in the Miles regime.
In addition to the large difference in the scales of the perturbation wavelength and wind velocity for the two modes, the growth rate of the KH mode is much larger than that for Miles' mode.
Furthermore, accounting for the surface tension effect  for prediction of the minimum wind speed is correct for Miles' regime, and
 has no physical meaning in the classical KH model.
Estimates (44) and (45) for the characteristic wave lengths in the Miles and KH regimes are consistent with respect to the orders of magnitude with those in Table I and Table II, respectively. In particular, the typical wave lengths in Miles' regime vary from $\lambda\sim 0.2m$ for $U_a=0.26 cm/s$ to $\lambda\sim 0.7cm$ for $U_a=0.38 cm/s$, while in the KH regime they range from $\lambda\sim 0.1m$  for $U_a=25m/s$  to  $\lambda\sim 10m$  for $U_a=50m/s$.

An important question is to what extent the PWL approximation provides a realistic representation of  smooth wind profiles for both the Miles as well as the KH regimes.

{\it Miles' regime.---}
A reasonable similarity with smooth wind profiles with regard to the stability characteristics should not be expected for the Miles modes in the two-layer PWL approximation of the wind velocity with wave lengths of the order the roughness scale, it rather models schematically the influence of the wind-shear effects on the interface instability.  However, even in such an approximation, the boundaries of the instability region in the Miles' regime are in fair quantitative agreement with more exact theories, and only the growth rate value is significantly overestimated. For small density ratio the specific feature of the PWL approximation in the Miles regime is the resonant nature of the solution that is characterized by a spectrum with a finite number of localized almost-discrete wave lengths. Indeed, numerical solutions of the two-layer PWL approximation for the wind speed demonstrate that the instability occurs in narrow intervals that are localized about either one (for $\sigma=0$) or two (for $\sigma>0$) resonant wave numbers, which are determined explicitly in the present modeling in the limit of small $\epsilon$. Furthermore, the number of resonant unstable waves grows with the number of the PWL sections $N$ such that the discrete spectrum tends to the continuous spectrum as the PWL profile tends to the continuous one in the limit of large $N$. The convergence rate strongly depends on the system parameters. In particular, numerical solutions of the Rayleigh equation with  $N$-layer PWL approximation for the wind profiles \cite{Gertsenshtein et al 1988}, in the Miles regime, indicate that the convergence in  $N$  is very slow for small density ratios $\epsilon^2\approx10^{-3}$, but quite good for larger values $\epsilon^2\approx10^{-1}$.

{\it KH-like regime.---}
\begin{figure}[!h]
\includegraphics[scale=0.45]{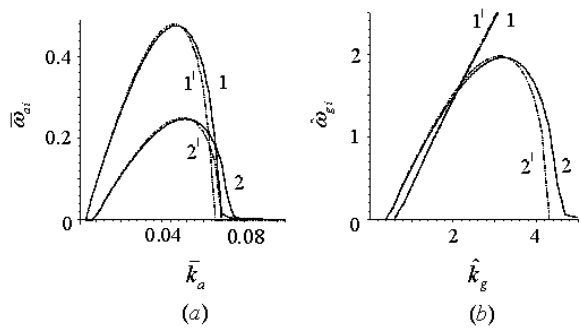}
\caption{\label{fig:epsfig4}
Growth rate vs wave number
in the plane (a) ($\bar{k}_{a}, \bar{\omega}_{ai}$) and (b) ($\hat{k}_{g}, \hat{\omega}_{gi}$);
curves 1, 1$'$ are for  $U_a=50m/s$, curves 2, 2$'$ are for $U_a=25m/s$  ($L_a=10^{-3}$).
Curves 1 and 2 are for tanh-profile of the wind
curves 1$'$ and 2$'$ are for the two-layer PWL approximation of the wind profile.
 }
 \end{figure}
Since the details of the wind profile are less essential when the wavelengths are much larger than the profile inhomogeneity length scale, it may be naturally expected that the two-layer PWL approximation  of the smooth profiles should be better within the moderately-short KH regime than for the short-waves Miles regime.  Numerical results for the smooth profile $\bar{U}(\bar{y})=tanh(\bar{y})$  have been obtained by solutions of the well-known Rayleigh equation for the inviscid flows (e.g. \cite{Alexakis et al 2002}). Indeed, unlike the behavior within  the Miles regime, the two-layer PWL approximation of the wind velocity for the KH regime provides a fair quantitative agreement for the growth rates with those for the smooth wind profile   (Fig. 4).

Figure 4 depicts the growth rate for the same parameters as Fig. 2c. As it is seen in Fig. 2c, the classical KH approximation of the growth rates is in a fair quantitative agreement with those in the two-layer PWL approximation (and hence with the results for $\bar{U}=tanh\bar{(y)}$) for sufficiently large wind speeds. The growth rates depicted in Fig. 4a in the characteristic scales in Eqs. (5) typical for Miles' region of parameters have been reconstructed in Fig. 4b in the characteristic scales  in Eqs. (6) natural for the KH regime. Note that if $\bar{k}_{a1}<\bar{k}_{a2}$  it may be that $\hat{k}_{g1}>\hat{k}_{g2}$  if
$k_1U_1^2>k_2U_2^2$ since $\hat{k}_{g1}=\epsilon^2 k_1U_1^2/g>\hat{k}_{g2}=\epsilon^2k_2U_2^2/g$ as it  occurs for the locations of the maximum growth-rates in Fig. 4.

\acknowledgments Helpful discussions with  A. Virtser
are gratefully acknowledged.


\renewcommand{\theequation}{A\arabic{equation}}
  \setcounter{equation}{0}  
  \section*{APPENDIX A. ASYMPTOTIC RELATIONS FOR MILES' REGIME}  

Omitting for simplicity subscript $a$ at all parameters and bar under $k$, $U$  and $\sigma$  in dispersion relation (13) yields
\begin{equation}
\Phi(\tilde{\omega})=-\epsilon^2\Psi(\tilde{\omega}),					
\end{equation}
where
\begin{eqnarray}
\Phi(\tilde{\omega})=\bigl [\tilde{\omega}-F(k)\bigr ]\Bigl [\tilde{\omega}^2-\frac{k+\sigma k^3}{U^2}\Bigr ],
\nonumber
\\
\Psi(\tilde{\omega})=\tilde{\omega}^3+\frac{k}{U^2}\tilde{\omega}
-F(k)\Bigl (\tilde{\omega}^2+\frac{k}{U^2}-\tilde{\omega}\Bigr )+\tilde{\omega}^2G(k).		
	 \nonumber
\end{eqnarray}
Equation (A1) leads to the following formal asymptotic expansion
\begin{equation}
\tilde{\omega}=\tilde{\omega}_0+\epsilon^2\tilde{\omega}_2+...
\end{equation}
Substituting (A2) into (A1) yields for the  first two terms in expansion (A2):
\begin{equation}
 \Phi(\tilde{\omega}_0)=0,\,\,\,\
 \tilde{\omega}_2=-\frac{\Psi(\tilde{\omega}_0)}{\partial \Phi(\tilde{\omega_0})/\partial \epsilon^2}.
\end{equation}
The formal solution (A3) is valid if
$\partial \Phi(\tilde{\omega_0})/\partial \epsilon^2\neq 0$. Otherwise (the resonance case) the asymptotic expansion (A2) should be replaced by:
\begin{equation}
\tilde{\omega}=\tilde{\omega}_0+\epsilon\tilde{\omega}_1+\epsilon^2\tilde{\omega}_2+...
\end{equation}
Substituting now expansion (A4) into (A1) yields the following new first two terms in expansion (A2):
 \begin{eqnarray}
 \Phi(\tilde{\omega}_0)=0,\,\,\,\
 \nonumber
\\
 \tilde{\omega}_1=i\sqrt{\frac{2\Psi(\tilde{\omega}_0)}{\partial^2 \Phi(\tilde{\omega}_0)/\partial \epsilon^2}}\equiv i\sqrt{\frac{F(k)[1+G(k)]}{2}},
\end{eqnarray}
since
 \begin{equation}
\Psi(\tilde{\omega}_0)=F^2(k)[1+G(k)],\,\,
\partial^2 \Phi(\tilde{\omega_0})/\partial \epsilon^2=4F(k).
 \nonumber
\\
\end{equation}
As a result, the unscaled value $\omega_1=\tilde{\omega_1}U$ with $U$  given by Eq. (17), is equal to
\begin{equation}
\omega_1= i\sqrt{\frac{1+G(k)}{2F(k)}(k+\sigma k^3)}.
\end{equation}

The asymptotic expansion (A4) fails for small $k\sim\epsilon$, when the leading order term in Eq. (A4), $\tilde{\omega}_0=F(k)\sim\epsilon^0 k^2$   becomes of the same order as the next term $\epsilon\tilde{\omega}_1\sim\epsilon k$  (here the estimates (4) for $F(k)$  and $G(k)$   have been used). Returning to the unscaled variables (12) and entire notations, this yields for the region of nonuniformity of Miles' mode:
\begin{equation}
\bar{k}\sim\epsilon,\,\,\,\bar{\omega}_0\sim \sqrt{\bar{k}}\sim \sqrt{\epsilon},
\,\,\epsilon\bar{\omega}_1\sim \epsilon/\sqrt{\bar{k}}\sim \sqrt{\epsilon}.
\end{equation}

\end{document}